\documentstyle[prl,aps]{revtex}
\begin{document}
\twocolumn[
\draft
\title{Non-Linear $\sigma$ Model for Inhomogeneous Spin Chains}
\author{Ken'ichi Takano \\
Laboratory of Theoretical Condensed Matter Physics and \\
Research Center for Advanced Photon Technology, \\
Toyota Technological Institute, Nagoya 468-8511, Japan}
%
\date{Recieved}
\maketitle
\widetext
\begin{quote}
      We derive a non-linear $\sigma$ model (NLSM) generally 
representing antiferromagnetic Heisenberg spin chains with 
inhomogeneous spin magnitudes and inhomogeneous 
nearest-neighbor exchange constants arrayed in finite periods. 
      Only a restriction is that the average of spin magnitudes on 
one sublattice is the same as that on the other sublattice. 
      The NLSM gives the gapless condition explicitly written 
by the spin magnitudes and the exchange constants.  
      We apply this gapless condition to several cases including 
systems with impurities. 
\end{quote}
]


\narrowtext

      In 1983 Haldane predicted that an antiferromagnetic Heisenberg 
spin chain is gapful if the spin magnitude is an integer, and is 
gapless if it is a half-odd-integer \cite{Haldane}. 
      Since his theory is based on mapping of a spin 
model to the non-linear $\sigma$ model (NLSM), the usefulness of 
the NLSM is recognized by researchers in condensed matter physics. 
      Various interesting aspects of the method of the NLSM are 
mentioned in some books \cite{Affleck,Fradkin,Tsvelik}. 
      Extension of the NLSM method to systems with inhomogeneity 
in spins and exchange constants is a challenging problem. 
      It is not only theoretically important but also useful in 
explaining and suggesting experiments. 
      Affleck reformulated the NLSM method and applied to a spin chain 
with bond alternation \cite{Affleck2}. 
      He divided the spin chain into blocks each of which 
contains two spins, and defined new variables for each block. 
      Affleck and Haldane analyzed and predicted gapless points for 
general homogeneous spin chains \cite{Affleck3}. 
      Fukui and Kawakami tried to extend the NLSM method to a spin 
chain including impurities and to spin chains with unit cell 
containing three and four spins \cite{Fukui0,Fukui1}. 
      They rather intuitively introduced new variables to form an NLSM 
and did not care to preserve the degrees of freedom. 
      Hence we cannot determine whether their resultant NLSM's are 
correct or not by inspecting their theory itself. 

      In this Letter, we obtain a NLSM generally representing 
a wide class of inhomogeneous antiferromagnetic spin chains. 
      The derivation is based on the idea dividing the system into 
blocks \cite{Affleck2} but blocks used here are generally defined 
to contain more than two spins. 
      In a path integral formula, 
the variables of integration for each block are transformed 
to form the NLSM with preserving the original degrees of freedom. 
      We apply this NLSM to several cases including  
systems with impurities and examine gapless conditions. 

      We consider the spin chain represented by the Hamiltonian 
\begin{equation}
      H = \sum_{j=1}^{N} J_j \, {\bf S}_j \cdot {\bf S}_{j+1}, 
\label{Hamiltonian}
\end{equation}
where ${\bf S}_j$ is the spin at site $j$ and $J_j (> 0)$ is the 
exchange constant between ${\bf S}_j$ and ${\bf S}_{j+1}$. 
      The number of lattice sites is $N$, the lattice spacing 
is $a$ and the system size is $L = a N$. 
      The quantum number for the magnitude of ${\bf S}_j$ is 
denoted by $s_j$. 
      The system is periodic with period $2b$ 
($b$: a positive integer): 
\begin{equation}
      J_{j+2b} = J_j , \quad s_{j+2b} = s_j . 
\label{period}
\end{equation}
     We divide the spin chain into blocks each of which contains 
$2b$ spin sites. 
      The block size $2ba$ is a positive integer (even number) times 
the size of the unit cell if a unit cell contains an even (odd) number 
of spins. 
      In a block $\{J_j\}$ and $\{s_j\}$ are arbitrary except for 
the restriction: 
\begin{equation}
      \sum_{j=1}^{b} s_{2j} = \sum_{j=1}^{b} s_{2j-1} . 
\label{restriction_s}
\end{equation}
      This restriction excludes systems with ferrimagnetic ground 
states in the classical limit. 

      The partition function $Z$ is written in a path integral formula. 
      When $(-1)^j s_j {\bf n}_j$ is the expectation value of 
${\bf S}_j$ for a coherent state, it is 
\begin{equation}
           Z = \int D[{\bf n}_j] \, \prod_j \delta({\bf n}^2_j - 1) \, e^{-S} . 
\label{partition}
\end{equation}
       The action $S$ at temperature $1/\beta$ is given by 
\begin{equation}
      S = i \sum_{j=1}^{N} (-1)^j s_j w[{\bf n}_j] 
+ \frac{1}{2} \int_0^{\beta} d\tau \sum_{j=1}^{N} 
                {\tilde J}_j \, (\delta {\bf n}_j)^2 
\label{action_n} 
\end{equation}
with $\delta {\bf n}_j = {\bf n}_{j+1} - {\bf n}_j$ and 
\begin{equation}
      \tilde J_j = J_j s_j s_{j+1} . 
\label{def_J_tilde}
\end{equation}
      The first term in Eq.~(\ref{action_n}) comes from the Berry 
phase and $w[{\bf n}_j]$ is the solid angle which the unit 
vector ${\bf n}_j$ forms in the period $\beta$. 

      We transform spin variables \{${\bf n}_{j}$\} into gradually 
changing variables \{${\bf m}(p)$\} and small fluctuations 
\{${\bf L}_q(p)$\}, 
where $p$ labels a block in the lattice $(p = 1, 2, \cdots , N/2b)$ 
and $q$ labels a site in the block $(q = 1, 2, \cdots , 2b)$. 
      The spin variable at the $q$th site in the $p$th block is written 
as 
\begin{eqnarray}
      {\bf n}_{2bp+q} &=& 
      \Bigl(1 - \gamma_q \frac{b-q}{2b} \Bigr) {\bf m}(p) 
       + \gamma_q \frac{b-q}{2b} {\bf m}(p-\gamma_q) \nonumber \\
                      & & + (-1)^q a {\bf L}_q(p) , 
\label{transform}
\end{eqnarray}
where $\gamma_q = +1 (-1)$ for $1 \le q \le b$ ($b+1 \le q \le 2b$). 
      The original constraints \{${\bf n}^2_j = 1$\} are changed to 
\{${\bf m}^2(p) = 1$\} and \{${\bf m}(p) \cdot {\bf L}_q(p) = 0$\}. 
      Here we notice that the number of variables increases in this 
transformation. 
      To solve the problem we can add an additional constraint 
for each block which is arbitrary as long as it consists with 
the other constraints; 
      e.~g. $\sum_{q=1}^{2b} (-1)^q {\bf L}_q(p) = 0$ 
or ${\bf L}_{2b}(p) = 0$ is a possible one. 
      Hence $2b$ vector variables 
$\{{\bf n}_{2bp+q} | \, q = 1, 2, \cdots , 2b\}$ 
per block are transformed to $2b$ vector variables 
$\{{\bf m}(p), {\bf L}_q(p) | \, q = 1, 2, \cdots , 2b-1\}$ per block; 
      ${\bf L}_{2b}(p)$ is written by the other variables through 
the additional constraint. 
      Also $2b$ original constraints 
$\{{\bf n}^2_q = 1 | \, q = 1, 2, \cdots , 2b\}$ per block 
are replaced by $2b$ constraints $\{{\bf m}^2(p) = 1, 
{\bf m}(p) \cdot {\bf L}_q(p) = 0 | \, q = 1, 2, \cdots , 2b-1\}$ 
per block. 
      Thus we obtained a new set of variables without changing 
the original degrees of freedom in the path integral formula 
(\ref{partition}). 
      In other words, we only transformed the variables of integration. 
      As for the additional constraint, the choice does not affect 
physical quantities at all. 
      This is because the variables 
$\{{\bf L}_q(p) | \, q = 1, 2, \cdots , 2b\}$ only appear as 
$\{{\bf L}_q(p)+{\bf L}_{q+1}(p) | \, q = 1, 2, \cdots , 2b-1\}$ 
in the partition function, as will be seen. 

      To take a continuum limit we identify the center of the $p$th 
block, $(2bp + b)a$, as coordinate $x$. 
      Then the difference between adjacent spin variables is 
replaced as  
\begin{eqnarray}
   \delta {\bf n}_{2bp+q} 
  \rightarrow a [ \partial_x {\bf m}(x) - {\bf R}_q(x) ] 
\label{delta_n}
\end{eqnarray}
with 
\begin{eqnarray}
   {\bf R}_q(x) = (-1)^q [{\bf L}_q(x) + {\bf L}_{q+1}(x)] . 
\label{delta_n_R}
\end{eqnarray}
Equation (\ref{delta_n_R}) for $q=2b$ reads as 
${\bf R}_{2b}(x) = {\bf L}_{2b}(x) + {\bf L}_{1}(x)$. 
      In the Berry phase term of the action (\ref{action_n}), the 
following relation stands due to the restriction 
(\ref{restriction_s}):  
\begin{equation}
      \sum_{q=1}^{2b} (-1)^q s_q w[{\bf n}_{2bp+q}] 
      = \sum_{q=1}^{2b} {\tilde s}_q \delta w[{\bf n}_{2bp+q}] , 
\label{relation_w}
\end{equation}
where $\delta w[{\bf n}_{2bp+q}] = 
w[{\bf n}_{2bp+q+1}] - w[{\bf n}_{2bp+q}]$ and
\begin{equation}
      {\tilde s}_q = \sum_{k=1}^{q} (-1)^{k+1} s_k . 
\label{def_s_tilde}
\end{equation}
     The continuum limit of the Berry phase term is taken 
after $\delta w[{\bf n}_{2bp+q}]$ is transformed to 
the $\tau$-integral of $\delta {\bf n}_{2bp+q} 
\cdot ({\bf n}_{2bp+q} \times \partial_{\tau} {\bf n}_{2bp+q})$ 
as in the usual way \cite{Fradkin,Tsvelik}. 
      Thus the action (\ref{action_n}) becomes $S_c = S_1 + S_2$ 
with 
\begin{eqnarray}
      S_1 &=& \int^{\beta}_0 d\tau \int^{L}_0 dx 
\biggl\{ - i \frac{s'}{2} 
(\partial_x {\bf m}) \cdot ({\bf m} \times \partial_{\tau} {\bf m}) 
\nonumber \\
& & + \frac{a}{2} {\bar J} (\partial_x {\bf m})^2 \biggl\} , \\ 
      S_2 &=& \int^{\beta}_0 d\tau \int^{L}_0 dx 
\frac{1}{2b} \sum_{q=1}^{2b} 
\biggl(
\frac{a}{2} {\tilde J}_q \/ {\bf R}_q^2 
+ i {\bf f}_q \cdot {\bf R}_q
\biggl) , 
\end{eqnarray}
where $s' = \sum_{q=1}^{2b} {\tilde s}_q / b$, 
${\bar J} = \sum_{q=1}^{2b} {\tilde J}_q / 2b$ and 
${\bf f}_q = i a {\tilde J}_q (\partial_x {\bf m})
+ {\tilde s}_q ({\bf m} \times \partial_{\tau} {\bf m})$. 

      The variables $\{{\bf L}_q\}$ appear only as $\{{\bf R}_q\}$ 
in the action $S_c$. 
      The variables $\{{\bf R}_q\}$ are not independent and 
equation $\sum_{q=1}^{2b} {\bf R}_q = 0$ stands due to the definition 
(\ref{delta_n_R}); e.~g. we can delete ${\bf R}_{2b}$ by the equation. 
      We treat the equation as a new constraint. 
      The constraints \{${\bf m} \cdot {\bf L}_q = 0$\} are 
rewritten as $\{{\bf m} \cdot {\bf R}_q = 0\}$.  
       Instead of deleting some variables by constraints, we 
insert the corresponding $\delta$-functions into the path integral 
formula (\ref{partition}) and treats all the variables independently. 
      We use the integral representations of 
$\delta(\sum_{q=1}^{2b} {\bf R}_q)$ and 
$\delta({\bf m} \cdot {\bf R}_q)$ 
with integration variables {\bf u} and $\alpha_q$. 
      Then the following term appears in addition to the action $S_c$: 
\begin{equation}
      S_3 = \int^{\beta}_0 d\tau \int^{L}_0 dx 
\frac{i}{2b} \sum_{q=1}^{2b} {\bf R}_q \cdot 
(- {\bf u} + \alpha_q {\bf m}) . 
\label{action_constraint}
\end{equation}
      Carrying out integrations in the partition function first 
with respect to \{${\bf R}_q$\} and then to {\bf u}, $S_2 + S_3$ 
reduces to 
\begin{equation}
      S'_2 = \int^{\beta}_0 d\tau \int^{L}_0 dx 
\frac{1}{4ba} \sum_{q=1}^{2b} \frac{1}{{\tilde J}_q} 
({\bf F}_q^2 - {\bf \bar F}^2) , 
\label{action_2_prime}
\end{equation}
where ${\bf F}_q = {\bf f}_q + \alpha_q {\bf m}$ and 
\begin{eqnarray}
    {\bf \bar F} 
= \sum_{q=1}^{2b} ({\tilde J}_q)^{-1} {\bf F}_q 
/ \sum_{q=1}^{2b} ({\tilde J}_q)^{-1} .  
\label{def_F}
\end{eqnarray}
      Expanding the integrand in Eq. (\ref{action_2_prime}), 
we find that \{$\alpha_q$\} appear only in a bilinear form 
and are integrated out. 

      Collecting $S_1$ and the remnant of $S'_2$ after the integration 
with respect to \{$\alpha_q$\}, we have the final effective action: 
\begin{eqnarray}
      &{}&  S_{\rm eff} = \int^{\beta}_0 d\tau \int^{L}_0 dx 
\biggl\{ 
- i \frac{J^{(0)}}{J^{(1)}} 
{\bf m} \cdot (\partial_{\tau} {\bf m} \times \partial_x {\bf m}) 
\nonumber \\ 
      &{}& + \frac{1}{2aJ^{(1)}} \biggl( 
\frac{J^{(1)}}{J^{(2)}} - \frac{J^{(0)}}{J^{(1)}} \biggl) 
(\partial_{\tau} {\bf m})^2 
+ \frac{a}{2} J^{(0)} (\partial_x {\bf m})^2 
\biggl\} , 
\label{action-final}
\end{eqnarray}
where \{$J^{(n)}$\} are defined as 
\begin{equation}
\frac{1}{J^{(n)}} = \frac{1}{2b} \sum_{q=1}^{2b} 
\frac{({\tilde s}_q)^n}{{\tilde J}_q}  \quad (n = 0, 1, 2) 
\label{def_J_n}
\end{equation}
with Eqs.~(\ref{def_J_tilde}) and (\ref{def_s_tilde}). 
      Thus we have obtained the action of the NLSM describing 
the Hamiltonian (\ref{Hamiltonian}) in the continuum limit. 
      The real-space cutoff is the length of a block of the minimum 
size. 
      The topological angle $\theta$ is given by setting $J^{(0)}/J^{(1)}$ 
in the first term as $\theta/4\pi$. 
      The velocity $v$ and the coupling constant $g$ are given by 
equating the coefficients of $(\partial_{\tau} {\bf m})^2$ and of 
$(\partial_{x} {\bf m})^2$ as $1/2gv$ and $v/2g$ respectively. 
      Note that the coefficient of $(\partial_{\tau} {\bf m})^2$ is 
always positive. 

      The action (\ref{action-final}) is independent of the way  
to divide the system into blocks. 
      First we displace each block by one site. 
      Then the spin magnitudes in a new block are ordered as 
$(s_2, s_3, \cdots , s_{2b}, s_1)$ instead of 
$(s_1, s_2, s_3, \cdots , s_{2b})$. 
      Denoting quantities related to the new blocks by letters with 
prime, we have relations ${\tilde s}'_q = s_1 - {\tilde s}_{q+1}$ 
for $1 \le q \le 2b-1$ and ${\tilde s}'_{2b} =0$. 
      Using these relations we obtain the following transformation: 
$J'^{(0)} = J^{(0)}$, 
\begin{eqnarray}
\frac{1}{J'^{(1)}} = \frac{s_1}{J^{(0)}} - \frac{1}{J^{(1)}} , \, 
      \frac{1}{J'^{(2)}} = 
\frac{s_1^2}{J^{(0)}} - \frac{2s_1}{J^{(1)}} + \frac{1}{J^{(2)}} . 
\label{displacement_J}
\end{eqnarray}
      This transformation do not change the coefficient of 
$(\partial_{\tau} {\bf m})^2$ in Eq.~(\ref{action-final}). 
      The topological angle (divided by $2\pi$) changes as 
\begin{eqnarray}
\frac{2J'^{(0)}}{J'^{(1)}} = 2 s_1 - \frac{2J^{(0)}}{J^{(1)}} . 
\label{transform_topology}
\end{eqnarray}
      Here the first term of $2s_1$ is an integer and does not 
affect physics. 
      The negative sign of the second term is also irrelevant. 
      Thus the action (\ref{action-final}) is invariant under the 
block displacement. 

      Second we inspect the effect for the action when we use 
the block of size $2rba$ ($r$: a positive integer) instead of $2ba$. 
      The order of the spin magnitudes in a new block is then $r$ 
times repetition of $(s_1, s_2, \cdots , s_{2b})$. 
      Because of the restriction (\ref{restriction_s}) we have relation 
${\tilde s}'_{2jb+q} = {\tilde s}_{q}$ for $q = 1, 2, \cdots, 2b$ and 
$j = 1, 2, \cdots, r$. 
      Using this relation we see that $J'^{(n)} = J^{(n)}$ for 
$n = 0, 1, 2$ and the action (\ref{action-final}) is invariant 
under the block enlargement. 

      The NLSM has a gapless excitation when $\theta/2\pi$ is 
a half-odd-integer. 
      This condition is written as 
\begin{equation}
\frac{2J^{(0)}}{J^{(1)}} = \frac{2l-1}{2} , 
\label{gapless_point}
\end{equation}
where $l$ is an arbitrary integer. 
      In what follows we examine 
this condition for several cases. 

      We first apply the general formula (\ref{action-final}) 
to the homogeneous case: $s_1 = s_2 = \cdots = s_{N} \equiv s$ 
and $J_1 = J_2 = \cdots = J_{N} \equiv J$. 
      In this case Eq.~(\ref{def_J_n}) gives $J^{(0)} = Js^2$, 
$J^{(1)} = 2Js$ and $J^{(2)} = 2J$. 
      The coefficients in Eq.~(\ref{action-final}) are 
simple as $\theta/4\pi = s/2$, $1/2gv = 1/8aJ$ and 
$v/2g = aJs^2/2$. 
      Hence Eq.~(\ref{action-final}) in this case is equivalent to 
the NLSM which Haldane originally considered 
\cite{Haldane}. 

      For $b=1$, a block contains only two spins and the 
restriction (\ref{restriction_s}) reads as $s_1 = s_2$. 
      Although the correct NLSM in this case has been already 
obtained \cite{Affleck,Affleck2}, we restate some 
results based on the general formula (\ref{action-final}). 
      \{$J^{(n)}$\} are calculated as 
$J^{(0)} = 2s^2 J_1 J_2 / (J_1+J_2)$, $J^{(1)} = 2sJ_1$ and 
$J^{(2)} = 2J_1$. 
     The gapless condition (\ref{gapless_point}) is then 
$s J_2/(J_1+J_2) = (2l-1)/4$. 
      This equation gives the single gapless point $J_2/J_1 = 1 (l=1)$ 
for $s = 1/2$ and $J_2/J_1 = 1/3 (l=1)$ for $s = 1$. 
      Numerical calculations for $s = 1$ show that the gapless point 
is at $J_2/J_1 = 0.6$ \cite{Kato,Yamamoto} and an experiment for 
[\{Ni(333-tet)($\mu$-N$_3$)\}$_{n}$](ClO$_4$)$_{n}$ 
agrees with this value \cite{Hagiwara}. 
      Hence the method of the NLSM does not always give 
quantitatively correct results. 
      However the NLSM is expected to represent 
the essence of quantum spin systems. 

      For $b=2$, the restriction (\ref{restriction_s}) is now 
$s_1 + s_3 = s_2 + s_4$. 
      In the case of $s_1 = s_2$, $s_3 = s_4$, $J_1 = J_3$ and 
$J_2 = J_4$, we have $2J^{(0)}/J^{(1)} = 2s_1s_3 (s_1+s_3) 
/ (s_1^2+s_3^2+2s_1s_3J_1/J_2)$. 
      Following the condition (\ref{gapless_point}), a gapless 
excitation appears at $J_2/J_1 = 4/7$ for $s_1 = 1/2$ and 
$s_3 = 1$, and at $J_2/J_1 = 3/7$ for $s_1 = 1/2$ and $s_3 = 3/2$. 
      For $s_1 = 1$ and $s_3 = 2$, two gapless points are 
$J_2/J_1 = 4/19$ and $4/3$. 
      Fukui and Kawakami \cite{Fukui1} have obtained the 
same results \cite{Fukui2}. 
      Tonegawa et al. performed numerical calculation for $s_1 = 1/2$ 
and $s_3 = 1$, and obtained $J_2/J_1 = 0.77 \pm 0.01$ for 
the gapless point \cite{Tonegawa}. 
      In the case of $s_1 = s_2$, $s_3 = s_4$, $J_1 = J_2$ and 
$J_3 = J_4$, we have $2J^{(0)}/J^{(1)} = 2s_1s_3/(s_1+s_3)$. 
      Hence the gapless excitation appears if and only if $s_1 = s_3$ 
and the value is a half-odd-integer irrespective of the values of 
$J_1$ and $J_3$. 
     In the case of $s_1 = s_2 = s_3 = s_4 \equiv s$, 
$J_1 = 1 - \delta$, $J_3 = 1 + \delta$ and $J_2 = J_4 \equiv J$, 
we have $2J^{(0)}/J^{(1)} = 2sJ/(J + 1 - \delta^2)$. 
      Chen and Hida performed numerical calculation for $s=1/2$ 
and obtained a phase boundary \cite{Chen}. 
      The positive-$J$ part of their boundary is close to 
$\delta = (1 - J)^{1/2}$ determined by Eq.~(\ref{gapless_point}) 
with $l=1$. 

      Using the block of $b=3$, we can deal with various systems. 
      Here we examine the case that a unit cell contains three sites. 
      In a block of the minimum size, the spin magnitudes and the 
exchange constants are ordered as $(s_1, s_2, s_3, s_1, s_2, s_3)$ 
and $(J_1, J_2, J_3, J_1, J_2, J_3)$. 
      The restriction (\ref{restriction_s}) is satisfied. 
      In this case we have $2J^{(0)}/J^{(1)} = s_1 - s_2 + s_3$ 
irrespective of the values of $J_1$, $J_2$ and $J_3$. 
      The condition (\ref{gapless_point}) says that systems with 
one or three half-odd-integer spins in a unit cell are gapless 
\cite{Fukui1,Fukui3}. 

      We examine a bond impurity system in the general formula 
(\ref{action-final}). 
      In this system all the spin magnitudes are the same and 
denoted by $s$; 
      $\{{\tilde s}_q\}$ are the same as those for the homogeneous 
case. 
      In contrast there are two kinds of exchange constants, 
$J_0$ and $J$. 
      We assume that $2h$ impurity bonds with $J_0$ randomly 
distribute among host bonds with $J$ in a block of size $2b$. 
      Then $(J^{(0)})^{-1}$ is calculated as $[1+\rho_0(J/J_0-1)]/Js^2$ 
with impurity density $\rho_0 = h/b$; 
      the randomness does not affect this quantity. 
      For $(J^{(1)})^{-1}$ we take an ensemble average since the 
contribution of an impurity bond changes whether the impurity site 
$q$ is even or odd. 
      We assume that all the possible distributions occur in the equal 
probability and that just half of the impurity bonds are on odd sites  
in the average. 
      Hence we have $(J^{(1)})^{-1} = [1+\rho_0(J/J_0-1)]/2Js$. 
      The topological angle is given by 
$\theta/2\pi = 2J^{(0)}/J^{(1)} = s$. 
      Therefore the bond impurities do not change the gapless 
condition of the homogeneous case \cite{smallJ0}. 
      Kawae et al. \cite{Kawae} argued that this model stands 
in the $s=1$ Haldane system (CH$_3$)$_4$NNi(NO)$_3$ (TMNIN) 
when nonmagnetic impurities Zn$^{2+}$ are doped. 
      They observed spin gaps at some impurity densities. 

      The last example is a site impurity system 
where impurity spins are located among host spins. 
      The spin magnitude is $s_0$ for an impurity spin against $s$ 
for a host spin. 
      Exchange constants are $J_0$ for both sides of an impurity 
site and are $J$ otherwise. 
      The number of impurity spins in a block is $2h$. 
      They are located on the sites of 
$\{k_j | \, j = 1, 2, \cdots , 2h\}$ in the block. 
      Their distribution is random in the block and the impurity 
density is $\rho_0 = h/b$. 
      In the calculation of ${\tilde s}_q$ on Eq.~(\ref{def_s_tilde}), 
we notice that it becomes 
${\tilde s}^{(0)}_q \equiv \sum_{k=1}^{q} (-1)^{k+1}s$ 
for a pure spin system. 
      We now replace the $k$th term in ${\tilde s}^{(0)}_q$ by 
$(-1)^{k+1} s_0$ if an impurity spin locates at the $k$th site. 
      Then we have 
\begin{equation}
      {\tilde s}_q = {\tilde s}^{(0)}_q 
      + (s-s_0) \sum_{j=1}^{2h} (-1)^{k_j} \theta(q - k_j) , 
\label{impurity_s_tilde}
\end{equation}
where the step function $\theta(x)$ is 1 for $x \ge 0$ and 0 
otherwise. 
      Taking the ensemble average with equal weight for all possible 
distributions, we have 
\begin{equation}
      \frac{2J^{(0)}}{J^{(1)}} = 
      s \frac{1+ (1 + \frac{s_0}{s}) {\tilde \rho_0}}
                 {1 + 2 {\tilde \rho_0}} 
\label{impurity_angle}
\end{equation}
with ${\tilde \rho_0} = \rho_0 (Js/J_0s_0 - 1)$. 
      Derivation of this equation will be reported elsewhere. 
      When $J_0s_0 < Js$, the gapless condition (\ref{gapless_point}) 
gives the following results. 
      In the case of $s=1$, Eq.~(\ref{gapless_point}) 
has no integer solution of $l$ for $s_0 \le 2$. 
      That is, impurities with $s_0 \le 2$ do not force the gapful 
excitation of a homogeneous $s=1$ spin system to be gapless 
\cite{smallJ0,Fukui4}. 
      In the case of $s=1/2$, Eq.~(\ref{gapless_point}) 
has no integer solution of $l$ for $1 \le s_0 \le 5/2$. 

      In summary, we obtained the NLSM (\ref{action-final}) for 
a general antiferromagnetic Heisenberg spin chain with 
inhomogeneous spin magnitudes and inhomogeneous 
nearest-neighbor exchange constants arrayed in a finite period. 
      We applied this formula to several cases and examined the 
gapless conditions. 
      Since the formula is general, it can be applied to various 
cases which were not treated here. 
      Extensions of the present NLSM method to ladder chains and 
two or three dimensional systems are future problems.



\begin{references}

\bibitem{Haldane} 
      F. D. M. Haldane, Phys. Lett. {\bf 93A}, 464 (1983); 
Phys. Rev. Lett. {\bf 50}, 1153 (1983). 

\bibitem{Affleck} 
      I. Affleck, 
"Field Theory Methods and Quantum Critical Phenomena" 
in {\it Fields, Strings and Critical Phenomena}, Les Houches 1988, 
eds. E. Brezin and J. Zinn-Justin, North-Holland, 1990. 

\bibitem{Fradkin} 
      E. Fradkin, {\it Field Theories of Condensed Matter Physics}, 
Addison-Wesley Publishing, 1994. 

\bibitem{Tsvelik} 
      A. M. Tsvelik, 
{\it Quantum Field Theories in Condensed Matter Physics}, 
Cambridge University Press, 1995. 

\bibitem{Affleck2} 
      I. Affleck, 
Nucl. Phys. B {\bf 257}, 397 (1985); {\bf 265}, 409 (1986). 

\bibitem{Affleck3} 
      I. Affleck F. D. M. Haldane, 
Phys. Rev. B {\bf 36}, 5291 (1987). 

\bibitem{Fukui0} 
      T. Fukui and N. Kawakami, Phys. Rev. {\bf B55}, 14709 (1997). 

\bibitem{Fukui1} 
      T. Fukui and N. Kawakami, cond-mat/9707148. 

\bibitem{Kato} 
      Y. Kato and A. Tanaka, J. Phys. Soc. Jpn. {\bf 63}, 1277 (1994). 

\bibitem{Yamamoto} 
      S. Yamamoto, J. Phys. Soc. Jpn. {\bf 63}, 4327 (1994). 

\bibitem{Hagiwara} 
      M. Hagiwara, Y. Narumi, K. Kindo, M. Kohno, H. Nakano, R. Sato, 
and M. Takahashi, Phys. Rev. Lett. {\bf 80}, 1312 (1998). 

\bibitem{Fukui2} 
      The case of $s_1 = s_2$, $s_3 = s_4$ and 
$J_2 = J_4 = 2 - (J_1 + J_3)/2$ was examined by another NLSM 
method in \cite{Fukui1}. 
      The major difference is in that the conservation of the 
degrees of freedom is not cared and the block is not introduced. 
      However their resultant action is the same as the action 
(\ref{action-final}) in this case. 

\bibitem{Tonegawa} 
      T. Tonegawa, T. Hikihara, T. Nishino, M. Kaburagi, S. Miyashita 
and H.-J. Mikeska, cond-mat/9712296. 

\bibitem{Chen} 
      W. Chen and K. Hida, private communication. 

\bibitem{Fukui3} 
      A further specialized case of $s_2 = s_3$ and 
$J_1 = J_2 = J_3$ was examined in \cite{Fukui1}. 
      The NLSM which they derived in another way does not consist 
with the present general action (\ref{action-final}). 
      However their action gives $\theta / 2 \pi = s_1$, which is 
the same as the present result for $s_2 = s_3$. 

\bibitem{smallJ0} 
      The NLSM method does not apply to the case of $J_0/J \ll 1$ 
both in the bond impurity system and the site impurity system. 
      The continuum approximation in the derivation of an NLSM 
is justified when adjacent spins are strongly coupled so that 
${\bf n}_j$ changes slowly as a function of $j$. 
      In the case of $J_0/J \ll 1$, however, local spin motion 
occurs about an impurity. 
      For the $s=1$ spin chain, an effective $s=1/2$ spin appears 
based on the valence-bond-solid picture: 
      I. Affleck, T. Kennedy, E. H. Lieb and H. Tasaki, Phys. Rev. Lett. 
{\bf 59}, 799 (1987). 
      Such $s=1/2$ spins are observed: 
      M. Hagiwara, K. Katsumata, I. Affleck, B. I. Halperin and 
J. P. Renard, Phys. Rev. Lett. {\bf 65}, 3181 (1990); 
      S. H. Glarum, S. Geshwind, K. M. Lee, M. L. Kaplan and J. Michel, 
Phys. Rev. Lett. {\bf 67}, 1614 (1991). 

\bibitem{Kawae} 
      T. Kawae, M. Ito, M. Mito, M. Hitaka and K. Takeda, 
J. Phys. Soc. Jpn. {\bf 66}, 1892 (1997). 

\bibitem{Fukui4} 
      Fukui and Kawakami examined a similar spin model but 
they considered impurity spins at regular intervals \cite{Fukui0}. 
      They derived another NLSM from the spin model in a different 
way. 
      Based on the NLSM they concluded that a gapful system can 
change to be gapless at some cases. 

\end{references}
\end{document}